\documentclass[useAMS,usenatbib]{mn2e}
\usepackage{graphicx}
 \usepackage{epstopdf}

\title[Pair creation in pulsars] {The novel mechanism of pair creation in pulsar magnetospheres}

\author[Osmanov et al.]{Z.N. Osmanov,$^{1,2}$\thanks{E-mail: z.osmanov@freeuni.edu.ge}
G.Z. Machabeli$^{3}$ \& N. Chkheidze$^{3}$\\
$^{1}$School of Physics, Free University of Tbilisi, 0183, Tbilisi, Georgia\\
$^2$E. Kharadze Georgian National Astrophysical Observatory, Abastumani 0301, Georgia\\
$^{3}$ ITP, Ilia State University, Tbilisi, Georgia\\
}

\begin{document}

\pagerange{\pageref{firstpage}--\pageref{lastpage}} \pubyear{2016}

\maketitle

\label{firstpage}

\begin{abstract}
In this paper we study the possibility of efficient pair production in a pulsar's magnetosphere. It has been shown that by means of the relativistic centrifugal force the electrostatic field exponentially amplifies. As a result the field approaches the Schwinger limit leading to pair creation process in the light cylinder area where the effects of rotation are very efficient. Analysing the parameters of the normal period ($\sim 1$ sec)  pulsars we have found that the process is so efficient that the number density of electron-positron pairs exceeds the Goldreich-Julian density by five orders of magnitude.

\end{abstract}

\begin{keywords}
pulsars: general - instabilities - acceleration of particles
\end{keywords}

\section{Introduction}

\cite{gold} has suggested that pulsars are rotating neutron stars centrifugally energising the pulsar's magnetospheric particles to energies enough for producing high energy electromagnetic radiation. Generally speaking, it is evident that the neutron stars rotate with increasing period of rotation, $P$, characterized by the slow down rate, $\dot{P}\equiv dP/dt>0$. The corresponding slow down power $\dot{W} = I\Omega|\dot{\Omega}|$, then becomes 
\begin{equation}
\label{power} 
\dot{W}\simeq 4.7\times 10^{31}\times P^{-3}\times\frac{\dot{P}}{10^{-15}}\times\frac{M_{ns}}{1.5M_{\odot}}\; ergs\; s^{-1},
\end{equation} 
where $\Omega$ is the angular velocity of rotation of the pulsar, the slow down rate is normalized by the parameters of the normal period pulsars (with period of the order of one second), $I = 2M_{ns}R_{\star}^2/5$ is the moment of inertia of the neutron star, $M_{ns}$ and $R_{\star}\simeq 10^6$cm denote its mass and radius respectively and $M_{\odot}\simeq 2\times 10^{33}$g is the solar mass. The value of  $\dot{W}$ indicates that a pulsar is a reservoir of enormous energy, which potentially might lead to very interesting processes.

The Gold's idea provoked a new theory of pulsar electrodynamics proposed by \cite{GJ}, where the authors have described the spacial distribution of charges inside the light cylinder (LC) surface (a hypothetical area where the linear velocity of rotation coincides with the speed of light) and studied efficiency of acceleration of particles escaping the inner magnetospheric area on open magnetic field lines. As it has been pointed out by \cite{sturrock}, the theory  of \cite{GJ} could not explain an emission pattern of radio pulsars. To study this particular problem \cite{sturrock} has examined a general case: when the dipolar momentum of a neutron star is inclined with respect to the rotation axis. The main idea is based on the fact that by means of interaction of very high energy (VHE) curvature gamma rays and strong magnetic field of the order of $10^{12}$ Gauss, the high efficiency pair production process initiates at the electron and proton polar zones. This model has been extended by \cite{jones}, where apart from the pair production by means of the curvature photons, the author also studied the mechanism of hadronic photo-absorption.  Several modifications of pair cascading have been proposed in a series of papers \citep{arons,DH,UM} considering the role of the strong magnetic field. 

In general, strong electric fields also can contribute to the efficient pair cascading process. In particular, according to electrodynamics, in vacuum the virtual particles and antiparticles are continuously producing and annihilating. If on the other hand, strong electric field is present, it can separate the virtual charges significantly, leading, via the quantum tunnelling, to pair creation. This happens when the work done on the Compton wave-length, $\lambda_{_C}$, equals the necessary energy of materialised pairs, $eE\lambda_{_C}\simeq 2mc^2$, where $m$ and $e$ are the electron's mass and charge respectively and $c$ is the speed of light. Original works dedicated to this problem has been considered in a series of papers \citep{sauter,heisen,schwinger}. The role of electric fields in pair cascading has been studied in the context of the relativistic ionic Coulomb fields \citep{CR} and an oscillating electric fields surrounding a pulsar \citep{PT}. In the latter, the authors considered the radial electric field induced by means of the radial currents and investigated the rate of pair production. It is worth noting that the pulsar current sheets provide induced electric fields small compared to the Schwinger limit $E_{_{S}} = \pi m^2c^3/(e\hbar)\simeq 1.4\times 10^{14}$ statvolt cm$^{-1}$ \citep{schwinger}, where $\hbar$ is the Planck constant. Therefore, it is clear that the aforementioned mechanism is not very efficient. On the other hand, in the pulsar magnetospheres the electric field might be induced by means of neutron star's rotation, ${\bf E} = {\bf \upsilon\times B}/c$, which is maximum close to the star's surface and for normal pulsars equals $2\times 10^8$ statvolt cm$^{-1}$ and still is by several orders of magnitude less than $E_{_{S}}$.

Since the pulsars are rapidly rotating neutron stars, it is clear that the centrifugal effects might be very important \cite{gold}. As it has been shown by \cite{mom} the centrifugal force in the LC area is different for different species of particles (magnetospheric electrons and positrons). This in turn, might lead to charge separation creating the Langmuir waves. On the other hand, since the centrifugal force is time dependent \citep{mom}, it acts as a parameter and the driven electrostatic field is parametrically amplified \citep{rep1,rep2,zev}. In the mentioned papers it was found that this mechanism is so efficient that the induced electric field exponentially increases, gradually approaches the Schwinger limit resulting in the efficient pair production. This is a completely new mechanism of pair creation in the pulsar magnetosphere and in this paper we study it.

The paper is organized as follows: in Sec.2 we introduce the theoretical model, in Sec. 3 we apply the approach to pulsars and obtain results and in Sec. 4 we sumarize them.

\section{Theoretical model}
In this section we present the theory of parametric instability of centrifugally excited Langmuir waves and apply the amplified electrostatic field to the pair production process. For this purpose, for the nearby zone of the LC we consider the linearized system of equations \citep{rep1,zev} composed by the Euler equation
 \begin{equation}
\label{eul1} \frac{\partial p_{\beta}}{\partial
t}+ik\upsilon_{\beta0}p_{\beta}=
\upsilon_{\beta0}\Omega^2r_{\beta}p_{\beta}+\frac{e_{\beta}}{m}E,
\end{equation}
the continuity equation
\begin{equation}
\label{cont1} \frac{\partial n_{\beta}}{\partial
t}+ik\upsilon_{{\beta}0}n_{\beta}, +
ikn_{{\beta}0}\upsilon_{\beta}=0
\end{equation}
and the Poisson equation
\begin{equation}
\label{pois1} ikE=4\pi\sum_{\beta}n_{\beta0}e_{\beta},
\end{equation}
where $\beta$ denotes the index of species (electrons and positrons), $p_{\beta}$ is the perturbation of momentum per unit mass, $r_{\beta}(t)\simeq \frac{V_{0\beta}}{\Omega_e}\sin\left(\Omega t+\phi_{\beta}\right)$, $\upsilon_{\beta}\simeq V_{0\beta}\sin\left(\Omega t+\phi_{\beta}\right)$ \citep{rep1} and $e_{\beta}$ are respectively the radial coordinate, radial velocity perturbation ($\upsilon_{\beta0}$ is the unperturbed value) and charge of the corresponding species, $V_{0\beta}$ is the velocity amplitude, $\phi_{\beta}$ denotes the phase, $E$ denotes the induced electrostatic field of the driven wave, $k$ - is its wave number and $n_{\beta}$ and $n_{\beta0}$ represent respectively perturbed and unperturbed particle number densities. Since the magnetic field in the pulsar's magnetosphere is very high, we assume that the particles are in the frozen-in condition ${\bf E_0 + \frac{1}{c}v_{0\beta}\times B_0 = 0}$ and are co-rotating with the field lines. The latter  are assumed to be almost straight, because it is believed that the emission is formed by the particles sliding along the open field lines.

Following the method developed in \citep{rep1,rep2} we introduce the anzatz
\begin{equation}
\label{ansatz}
n_{\beta}=N_{\beta}e^{-\frac{iV_{0\beta}k}{\Omega}\sin\left(\Omega t
+ \phi_{\beta}\right)},
\end{equation}
which reduces the aforementioned set of equations to non-autonomous "Mode" equations for electrons and positrons
\begin{equation}
\label{ME1} \frac{d^2N_e}{dt^2}+{\omega_e}^2 N_e= -{\omega_e}^2 N_p
e^{-i \chi},
\end{equation}
\begin{equation}
\label{ME2} \frac{d^2N_p}{dt^2}+{\omega_p}^2 N_p= -{\omega_p}^2 N_e
e^{i \chi},
\end{equation}
where $\omega_{e,p}\equiv\sqrt{4\pi e^2n_{e,p}/m\gamma_{e,p}^3}$ and
$\gamma_{e,p}$ are the relativistic plasma frequencies and the
relativistic factors for the beam components and $\chi = b\cos\left(\Omega t+\phi_{+}\right)$, $b =
\frac{2ck}{\Omega_e}\sin\phi_{-}$, $2\phi_{\pm} = \phi_p\pm\phi_e$. 
After making the Fourier transform of Eqs. (\ref{ME1},\ref{ME2}) one straightforwardly derives the dispersion relation of the Langmuir wave

\begin{equation}
\label{disp} 
\omega^2-\omega_e^2-\omega_p^2\;J^2_0(b) =\omega^2\omega_p^2\sum_{\mu\neq 0}\frac{J^2_{\mu}(b)}{\left(\omega+\mu\Omega\right)^2}.
\end{equation} 

To study the instability the frequency must be decomposed by the real and imaginary parts $\omega = \omega_r+\Delta$. The former should satisfy the resonance condition $\omega_r = \mu_{r}\Omega_e$ for the most efficient process \citep{rep1,zev,rep2}. Then, one obtains the following cubic equation
 \begin{equation}
 \label{disp1}
 \Delta^3=\frac{\omega_r {\omega_p}^2 {J_{\mu_{r}}(b)}^2}{2},
 \end{equation}
having two complex solutions characterising a growth rate of the centrifugally excited Langmuir waves
\begin{equation}
 \label{grow}
 \Gamma= \frac{\sqrt3}{2}\left (\frac{\omega_e {\omega_p}^2}{2}\right)^{\frac{1}{3}}
 {J_{\mu_{r}}(b)}^{\frac{2}{3}},
\end{equation}
where $\omega_r = \omega_e$ and $J_{\mu_{r}}$ denotes the Bessel function of the first kind.

Therefore the time evolution of the electrostatic field is given as follows
\begin{equation}
 \label{Et}
 E(t) = E_0e^{\Gamma t},
\end{equation}
where $E_0$ is its initial value which can be approximated straightforwardly
\begin{equation}
 \label{E0}
 E_0 \simeq 4\pi n_{_{GJ}}e\Delta r,
\end{equation}
where 
\begin{equation}
 \label{GJ}
 n_{_{GJ}} = \frac{\Omega B}{2\pi ec}\times\frac{1}{1-r^2/R_{lc}^2},
\end{equation}
is the Goldreich-Julian number density \citep{GJ}, $R_{lc} = c/\Omega$ denotes the LC radius and $\Delta r$ is the characteristic length-scale of charge separation, which takes place close to the LC area. Its value can be derived from the fact that the process of CA occurs in the LC zone, where the Lorentz factor asymptotically increases \citep{RM}
\begin{equation}
\label{gam} 
\gamma = \frac{\gamma_0}{(1-r^2/R_{lc}^2)},
\end{equation} 
therefore, $\Delta r$ should characterise a scale factor of non-uniformity, which can be expressed as $\Delta r = \frac{\gamma}{\partial\gamma/\partial r}$, finally leading to $\Delta r\simeq \gamma_0R_{lc}/(2\gamma)$ \citep{zev}. Here $\gamma_0$ is the initial relativistic factor. 

As it turns out from Eq. (\ref{Et}) the field is exponentially increasing and in due course of time the electric field will inevitably reach the Schwinger limit and thus, the probability of pair production might increase very rapidly. 

In general for a constant electric field the pair production rate ($R$ - number of produced pairs per unit of time and unit of volume) is given by \citep{schwinger,MP}
\begin{equation}
 \label{rate}
 R\equiv\frac{dN}{dtdV} = \frac{e^2E^2}{4\pi^3c\hbar^2}\sum_{n}\frac{1}{n^2}exp\left({-\frac{\pi m^2c^3}{e\hbar E}n}\right).
\end{equation}
Generally speaking, if the electrostatic field is varying the pair production rate is higher, but in the considered scenario the plasma oscillation frequency can be maximum $10^{4-5}$ Hz, which is by many orders of magnitude less than $\nu\simeq 2mc^2/h\sim 10^{20}$ Hz. Therefore, the constant field case is a physically realistic approximations and hence, the aforementioned formula is valid.

\section{Discusion}

\begin{figure}
  \resizebox{\hsize}{!}{\includegraphics[angle=0]{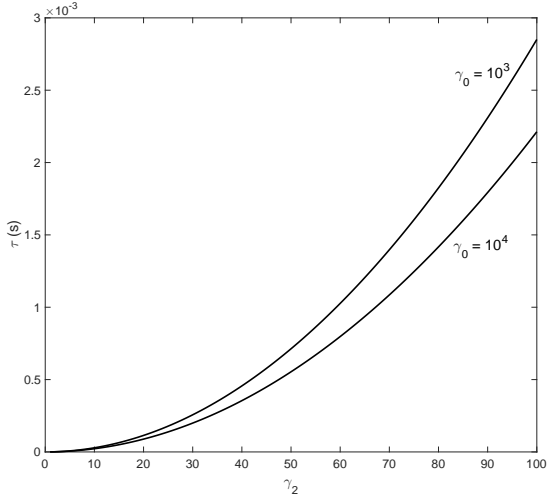}}
  \caption{Behaviour of the instability time-scale versus $\gamma_2$.
The set of parameters is: $P = 1$ sec, $\dot{P} = 10^{-15}$ s s$^{-1}$, $\gamma_1 = 10^3$, $\gamma_0 = \{10^3; 10^4\}$.}\label{fig1}
\end{figure}

\begin{figure}
\resizebox{\hsize}{!}{\includegraphics{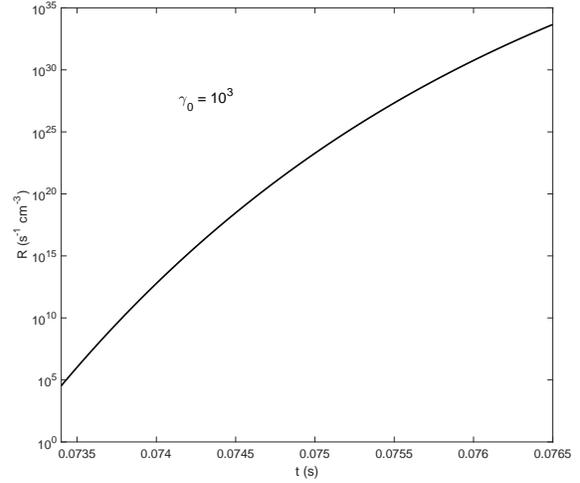}}
\resizebox{\hsize}{!}{\includegraphics{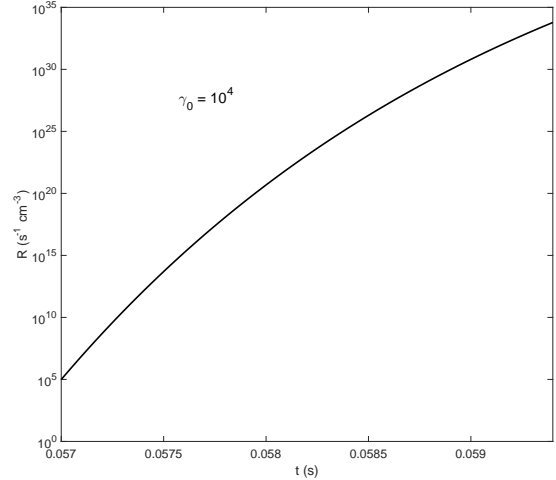}}
 \caption{Two plots for the behaviour of the pair production rate versus time are shown. The set of parameters is the same as in Fig. 1. On the top panel we show the plot for $\gamma_0 = 10^3$ and on the bottom panel we show the results for $\gamma_0 = 10^4$.}
 \label{fig2}
\end{figure}

\begin{figure}
  \resizebox{\hsize}{!}{\includegraphics[angle=0]{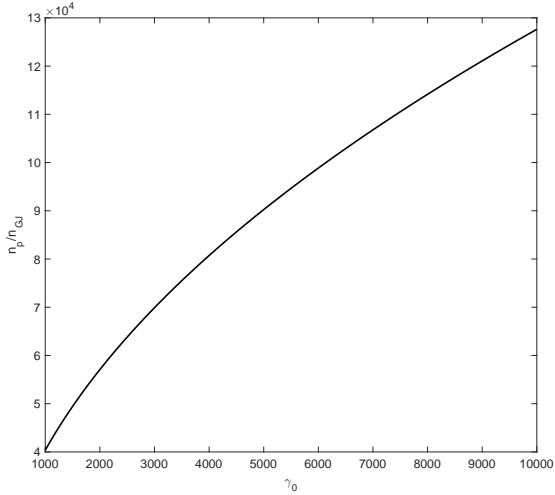}}
  \caption{Dependence of the multiplicity factor versus the initial Lorentz factor.
The set of parameters is the same as in Fig. 2 except $\gamma_0$ which changes in the interval $[10^3-10^4]$.}\label{fig1}
\end{figure}

In this section we consider the normal period pulsars and study the possibility of pair production by means of the centrifugally driven electrostatic fields. On the LC length-scale, where the CA is most efficient the magnetic induction \citep{MT}
\begin{equation}
 \label{B}
 B_{st}\simeq 10^{12}\times\left(\frac{P}{1\; s}\times\frac{\dot{P}}{10^{-15}\; ss^{-1}}\right)^{1/2} \;\;Gauss,
\end{equation}
guarantees frozen-in condition, leading to a one-dimensional geometry of particles' kinematics along the magnetic field lines.

Using the fact that CA lasts until the plasma energy density, $\gamma mc^2 n_{_{GJ}}$, exceeds that of magnetic field, $B^2/(8\pi)$, for the maximum Lorentz factor one obtains

\begin{equation}
\label{gmax} 
\gamma_{m}\simeq 2.6\times 10^5\times P^{5/4}\times\left(\frac{\gamma_0}{10^4}\right)^{1/2}\times\left(\frac{\dot{P}}{10^{-15}s\; s^{-1}}\right)^{1/4},
\end{equation} 
where we have taken into account Eq. (\ref{gam}), the radial behaviour of the dipolar magnetic field and the corresponding expression on the LC, $B\simeq B_{st} R^3_{st}/R_{lc}^3$ . Here $R_{st}\simeq 10$km denotes the neutron star's radius. 

To demonstrate efficiency of the Langmuir wave's amplification, we consider two different initial Lorentz factors $\gamma_0 = \{10^3; 10^4\}$. In Fig. 1 we plot the dependence of the instability time-scale on $\gamma_2$. The set of the rest of parameters is: $P = 1$ sec, $\dot{P} = 10^{-15}$ s s$^{-1}$, $\gamma_1 = 10^3$, $R_{st} = 10$ km. In the framework of this paper we assume the equipartition approximation: when energy is uniformly distributed among all species: $\gamma_{m}n_{_{GJ}}\simeq n_1\gamma_1\simeq n_2\gamma_2$. As it is clear from the plots the instability time-scales are small compared to the escape time-scale (which is of the order of the rotation period of the pulsar $P\simeq 1$ sec), indicating high efficiency of the amplification process. 

Generally speaking, the proposed mechanism should be somehow terminated. If the co-rotation is maintained this will happen when the produced total energy density of electron positron pairs equals the energy density of the electrostatic field, leading to the following integral expression
\begin{equation}
\label{term1} 
\sum_{n}\frac{1}{n^2}\int_0^{\tau}\exp\left(-\frac{\pi m^2c^3n}{e\hbar E_0}e^{-\Gamma t}\right)dt = \frac{\pi^2\hbar^2}{4mce^2},
\end{equation} 
which actually is an equation for determining the value of $\tau$ when the pair production is terminated. After integration one can straightforwardly derive the corresponding algebraic equation
$$\sum_n\frac{1}{n^2}\left[E_1\left(-\frac{\pi m^2c^3n}{e\hbar E_0}e^{-\Gamma\tau}\right)-E_1\left(-\frac{\pi m^2c^3n}{e\hbar E_0}\right)\right] =$$
\begin{equation}
\label{term2} 
= \frac{\pi^2\hbar^2}{4mce^2}\Gamma,
\end{equation} 
where $E_1$ denotes the exponential integral.

For the same set of parameters as in Fig. 1 one can solve Eq. (\ref{term2}) for $\tau$. In particular, for $\gamma_0 = 10^3$ and $\gamma_0 = 10^4$ the corresponding termination time-scales equal $\sim 7.6\times 10^{-2}$ s and $\sim 5.9\times 10^{-2}$ s respectively. 

By taking these values into account one can estimate the pair production rate from Eq. (\ref{rate}). On Fig. 2 we plot the dependence of $R$ versus time. The set of parameters is the same as in Fig. 1. As it is clear from the plots the pair creation rate is a continuously increasing function of time, which is a natural result of the fact that by means of the parametric instability, the electrostatic field exponentially amplifies (see Eq. (\ref{Et})). One has to note that the absolute values of $R$ are unrealistically high, indicating that some additional mechanism of termination of the process must exist.

The proposed mechanism strongly depends on the effects of relativistic centrifugal force, which takes place only if the energy density of particles is less than that of the magnetic field. But in due course of time the number density of magnetospheric electron-positron pairs, $n_p$, increases very rapidly and as soon as the following condition  
\begin{equation}
\label{cond} 
2\gamma_{rot}mc^2n_p<\frac{B^2}{8\pi}
\end{equation} 
violates, the co-rotation stops and consequently the process is terminated. Here $\gamma_{rot} = (1-r^2/R_{lc}^2)^{-1/2}$ is the Lorentz factor corresponding to rotation. After taking into account Eq. (\ref{gam}) one can straightforwardly derive the maximum number density of pairs, $n_p$ on the LC zone
$$n_p\simeq \frac{B^{7/4}}{16\pi mc^2}\times\left(\frac{4mc\gamma_0\Omega}{e}\right)^{1/4}\simeq$$
\begin{equation}
\label{np} 
\simeq 4.17\times 10^5\times P^{-37/8}\times\left(\frac{\dot{P}}{10^{-15}}\right)^{7/8}\times\left(\frac{\gamma_0}{10^4}\right)^{1/4}.
\end{equation} 
On Fig. 3 we plot the behaviour of the multiplicity factor $n_p/n_{GJ}$ versus the initial relativistic factor. The set of parameters is the same as in Fig. 2 except $\gamma_0$ which varies in the interval $[10^3-10^4]$. As one can see from the plot the multiplicity factor is a continuously increasing function of the initial Lorentz factor and for $10^4$ reaches the value of the order of $1.3\times 10^5$ indicating high efficiency of the process.

\section{Summary}

We have examined centrifugally accelerated pulsar's magnetospheric particles and studied the parametric amplification of the electrostatic field. 

Considering the typical pulsar parameters it has been shown that the electric field exponentially increases and gradually reaches the Schwinger limit, when efficient pair creation might occur.

Analysing constraints imposed on the process we have found that the mechanism terminates when the energy density of produced particles exceeds that of the magnetic field. As a result, it has been shown that this process leads to the multiplicity factor compared to the GJ number density of the order of $10^5$.

Given that the novel mechanism of pair production substantially changes the number density of electron-positron plasmas in pulsar's magnetospheres it might significantly influence the physical processes there. In particular, it is evident that the processes of particle acceleration strongly depends on the plasma density. Consequently, the emission spectral pattern will be influenced as well by the efficient pair creation. This in turn, might give rise to coherent radio emission, which seems to be quite promising in the light of modern enigma - fast radio bursts. Since all these problems are beyond the intended scope of the paper  we are going to consider them very soon.


\section*{Data Availability}

Data are available in the article and can be accessed via a DOI link.

\end{document}